\documentclass{article}
\usepackage{mrs2005,epsfig}

\newbox\grsign      \setbox\grsign=\hbox{$>$}
\newdimen\grdimen   \grdimen=\ht\grsign
\newbox\simgreatbox \setbox\simgreatbox=\hbox{\raise.5ex\hbox{$>$}\llap
                        {\lower.5ex\hbox{$\sim$}}}\ht1=\grdimen\dp1=0pt
\newbox\simlessbox  \setbox\simlessbox =\hbox{\raise.5ex\hbox{$<$}\llap
                        {\lower.5ex\hbox{$\sim$}}}\ht2=\grdimen\dp2=0pt
\def\simgreat{\mathrel{\copy\simgreatbox}}
\def\simless {\mathrel{\copy\simlessbox }}
 
\begin{document} 
\title{CLUSTER EVOLUTION SINCE $z \sim 1$}

\author{Ricardo Demarco, John P. Blakeslee, Holland C. Ford, Nicole L. Homeier, \\ Simona Mei and Andrew W. Zirm}
\affil{Dept. of Physics \& Astronomy, Johns Hopkins University \\ 3400 N. Charles Street. Baltimore, MD 21218, USA} 
 
\author{Piero Rosati and Alessandro Rettura}
\affil{European Southern Observatory \\ Karl-Schwarzschild-Str. 2, D-85748, Garching bei M\"unchen, Germany} 

\author{Chris Lidman}
\affil{European Southern Observatory \\ Alonso de Cordova 3107, Vitacura, Santiago, Chile} 

\author{Roderik A. Overzier and George Miley}
\affil{Leiden Observatory \\ Postbus 9513, 2300 RA Leiden, The Netherlands} 
 
\begin{abstract} 
We have used the Advanced Camera for Surveys (ACS) on the HST to
obtain optical imaging of a sample of 8 galaxy clusters in the
redshift range 0.8 $< z <$ 1.3. The ACS data provides accurate
photometry and the ACS high angular resolution makes it possible to
obtain crucial morphological information of the cluster galaxy
populations. These data are supported by X-ray observations and by
ground based imaging and multi-object spectroscopy. We will present
results from this multi-wavelength study which is providing a
comprehensive view of the evolution of structures in the universe
since $z \sim 1$. The structure of the ICM, galaxy and DM components
of the clusters are presented and compared. The spectrophotometric
properties of cluster galaxies are investigated in the context of
galaxy evolution and the formation epoch of massive early-type
galaxies in these clusters is estimated to be at $z > 2$. We will also
present results from our study of the evolution of the
Morphology-Density relation in clusters since $z \sim 1$ down to the
present day universe. Our cluster evolution studies are also being
complemented by observations of protoclusters at redshift $z \simgreat
2$.
\end{abstract} 
 
\section{Introduction} 

A major fraction of the galaxies in the present universe are found
isolated in the field. Only about 5\% of the galaxies in the local
universe belong to gravitationally bound structures such as small
groups and clusters of galaxies. Cluster and field environments can
affect galaxies in different ways, therefore, a detailed studied of
galaxy properties in clusters and in the field is crucial to
understand galaxy evolution.

In this respect, clusters of galaxies are ideal laboratories to
investigate the influence of high density regions on galaxy properties
such as morphology, color and star-forming activity. In addition, the
study of the structure of the different components of clusters,
baryons and dark matter (DM), allows us to characterize the dynamical
state of the cluster and to better understand the assemblage of large
scale structures in the universe. In massive clusters, baryons
(galaxies and gas) accumulate in the potential well created by the
cluster DM halo. In such an environment, galaxy properties can be
affected by interactions between galaxies and the surrounding
intracluster medium (ICM), however the details of the physical
mechanisms operating in such interactions are still not well
understood.

Galaxy clusters are hosts of the most massive and oldest early-type
galaxies known in the universe. Determining their epoch of formation
and understanding their evolution, as well as that of the other
cluster galaxy populations, is one of the major tasks in modern
cosmology. Such an evolutionary study demands the study of galaxy
properties up to redshifts $z \sim 1$. The well defined
color-magnitude relation (CMR) in clusters has been investigated to
estimate the age and mode of formation of massive cluster early-type
galaxies up to $z \sim 1$ \cite{b03,lrd04}.  The results drawn from
these studies indicate that massive cluster elliptical galaxies are
formed at high redshift ($z > 2$), in agreement with studies of red
objects in the field at $z > 1$ \cite{dcr00,d03,p03}.

However, the major problem in any study of galaxy properties in high
density environments at high redshift is to actually find those high
density regions. To date, only a few high redshift clusters at $z
\simgreat 1$ have been spectroscopically confirmed (see \cite{mrl05},
and references therein). Beyond $z = 1.4$ and below $z \sim 2$ there
are no confirmed galaxy clusters. This redshift interval is of great
importance since it is in this epoch when aggregations of matter are
expected to start virializing to form X-ray luminous ICMs. Some of
these regions could be traced by galaxy overdensities physically
associated to radio galaxies \cite{blm03} which are formed at earlier
stages in cosmic time.  There is clear evidence of the existence of
such early structures, the so called proto-clusters
\cite{v02,m04,sas05,o05}. These structures may thus be the progenitors
of the massive galaxy clusters we observe in the local
universe. Hence, the study of the physical properties of galaxies in
these structures should give us a good idea of how galaxies form and
evolve in the peaks of the cosmic matter distribution. In particular,
some of those galaxies, including radio galaxies, may be the
progenitors of massive cluster early-type galaxies \cite{z05}.

\section{The ACS cluster program} 

In order to address the above issues, in particular the formation of
massive early-type galaxies and the effects of the environment on the
galaxy populations, we have carried out an ambitious program with the
Advanced Camera for Surveys (ACS; \cite{f98}) to study galaxy
properties in clusters and proto-clusters at 50\% of the current age
of the universe. The sample of galaxy clusters in our ACS program is
composed of 8 rich clusters in the range 0.8 $< z <$ 1.3
(corresponding to a lookback time between 7 and 9 Gyr). The
proto-clusters under study are galaxy overdensities associated with 3
different powerful radio galaxies at 2.2 $< z <$ 5.2. Moreover, ACS
has been used to observe galaxy clusters with strong lensing features
at lower redshift \cite{b05} and has also allowed us to discover
gravitational arcs in clusters at $z \simgreat 1$. Together with weak
lensing studies from ACS data \cite{jwb05,l05}, these strong lensing
investigations with the ACS are opening the way to map with high
accuracy the mass distribution of some of the most distant clusters of
galaxies known to date.

The ACS observations were carried out with the Wide Field Camera (WFC)
and included imaging in 2 to 3 different bands with the filters of the
Sloan Digital Sky Survey (r [F625W], i [F775W] and z [F850LP]). The
high angular resolution and quantum efficiency of the ACS make it
possible to obtain accurate photometry and unprecedented morphological
details of galaxies at redshifts as high as $z \sim 1.3$. The ACS WFC
delivers a field of view of 202\arcsec\ $\times$ 202\arcsec\ with a
pixel scale of 0.049\arcsec. In several cases, there are multiple
pointings available per cluster, with all pointings overlapping the
central 1\arcmin\ cluster region.

In addition to the ACS guaranteed time observation (GTO) program, a
multi-wavelength data set, from X-rays to infrared (IR), including
imaging and spectroscopy, has been built to support the ACS data and
to provide one of the most complete and comprehensive studies of
distant clusters and proto-clusters so far. Space based observations
were carried out with Chandra (imaging), Newton-XMM (imaging and
spectroscopy) and Spitzer (imaging) while the ground based data were
collected with the ESO NTT (imaging), ESO VLT (imaging and
spectroscopy) and Keck (imaging and spectroscopy).

\begin{figure}
\vspace*{0.25cm}
\begin{center}
\epsfig{figure=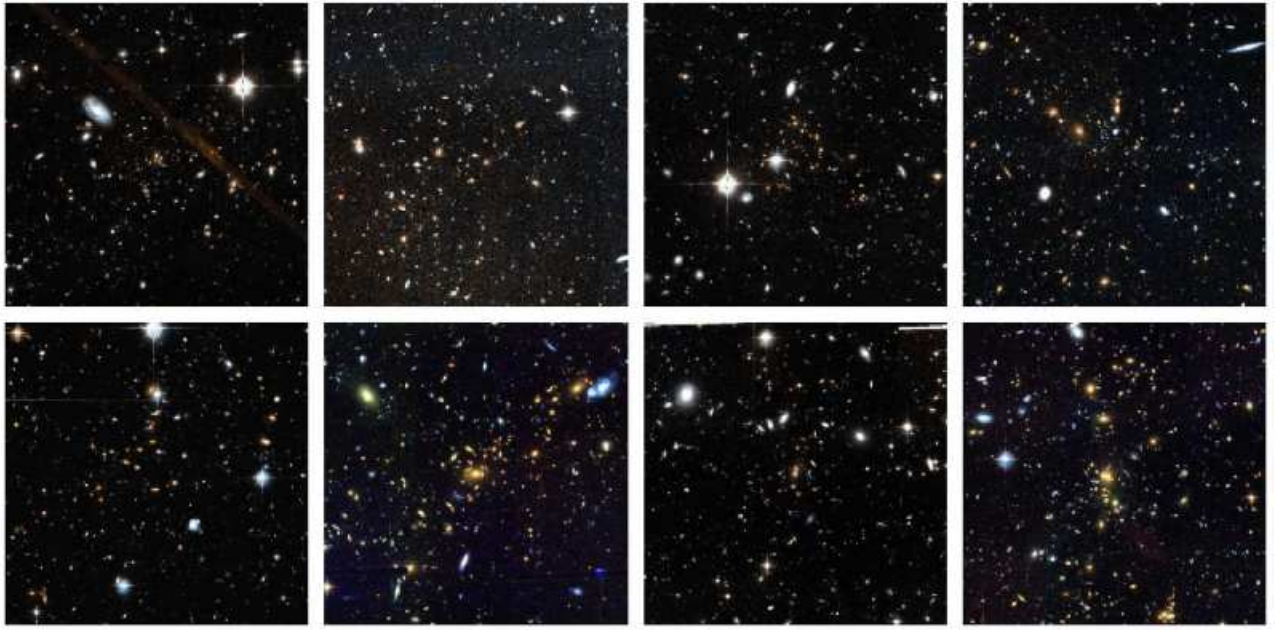,width=17cm}
\end{center}
\vspace*{0.25cm}
\caption{Intermediate redshift clusters observed with the ACS. Top row
from left to right: RXJ0848+4452 ($z = 1.26$), RXJ0848+4453 ($z =
1.27$), RXJ0910+5422 ($z = 1.10$) and CL1604+4304 ($z = 0.90$). Bottom
row from left to right: CL1604+4321 ($z = 0.92$), MS1054-03 ($z =
0.83$), RDCSJ1252-2927 ($z = 1.24$) and RXJ0152-1357 ($z = 0.84$).
All cutouts are 2\arcmin\ on a side. \label{acs_cls}}
\end{figure}

A summary of the ACS data available on the eight galaxy clusters is
presented in Table \ref{tab_cls}. In Figure \ref{acs_cls} we show ACS
color images of the observed intermediate redshift clusters. A more
detailed description of the ACS, X-ray and Spectroscopic data is
presented in Postman et al. (2005; \cite{p05}).

\begin{center}
\begin{table}
\caption{ACS intermediate redshift cluster sample. \label{tab_cls}}
\begin{tabular}{lccclc}
\hline
Cluster           & Redshift & $\sigma_v (km/s)$ & $L_x$ (10$^{44}$ erg/s) & ACS filters               & Total HST orbits \\
\hline
\hline
MS1054-03         & 0.83     & 1112              & 23.3                    & V, {\it i}, {\it z}       & 20 \\
RXJ0152-1357      & 0.84     & 1300              & 7.8                     & {\it r}, {\it i}, {\it z} & 40 \\
CL1604+4304       & 0.90     & 1200              & 2.0                     & V, I                      & 4  \\
CL1604+4321       & 0.92     & 935               & $<$1.2                  & V, I                      & 4  \\
RXJ0910+5422      & 1.10     & -                 & 1.5                     & {\it i}, {\it z}          & 8  \\
RDCSJ1252-2927    & 1.24     & 755               & 2.5                     & {\it i}, {\it z}          & 32 \\
RXJ0848+4452      & 1.26     & -                 & 1.0                     & {\it i}, {\it z}          & 24 \\
RXJ0848+4453      & 1.27     & 640               & 1.5                     & {\it i}, {\it z}          & 24 \\
\hline
\end{tabular}
\end{table}
\end{center}

\section{Intermediate redshift clusters of galaxies} 

The large amount and high quality of data gathered during the ACS
cluster program makes it possible to study in detail the assembly and
evolution of structures in the universe, from cluster scale down to
the scale of massive cluster early-type galaxies, during the last 8.7
Gyrs of cosmic history. Extensive redshift surveys of some of the ACS
clusters (e.g., \cite{tkvd99,vdff00,drl05}) combined with the
available X-ray, optical and near-IR data show clear evidence of
massive clusters being assembled through major mergers of smaller
clusters or groups at $z \sim 0.8$ (e.g., \cite{jwb05,gdr05}). At
redshift greater than unity, clusters show disturbed morphologies in
their gas, galaxy and DM distribution (e.g., \cite{l05}; Mei et al.,
ApJ, submitted) suggestive of large scale mergers occuring at a large
lookback time. Most of distant clusters would thus be in an
unvirialized state, being formed in a hierarchical way through the
accretion of smaller units which may also be part of larger scale
filaments (\cite{k05};Kodama et al., in prep.). Gas and galaxies,
trapped in the potential of DM haloes, follow the dynamics of the
merger (with the gas being affected by ram pressure) tracing the
spatial distribution of the forming massive structure
\cite{r04,jwb05,jwf05}. These mergers can continue down to the local
universe, as shown by observations of lower redshift and nearby
clusters (e.g., \cite{hm96,blc02,mgd02}).

Identification of the passive and star-forming cluster galaxy
populations at $z \sim 1$ shows that the high density regions of
clusters are dominated by passive (non star-forming) galaxies, while
the star-forming galaxies mostly populate the outer cluster
regions. This result, in agreement with the Butcher-Oemler effect,
extend to this early epoch the evidence of the existence of such a
segregation previously observed in lower redshift clusters (e.g.,
\cite{ely01}). In the cluster RXJ0152 ($z = 0.84$) the star-forming
galaxy population avoids regions where the X-ray emission is greater
than the 3$\sigma$ level \cite{drl05}. These observations are
consistent with the simple view in which galaxies falling into the
central cluster regions get their star-formation activity suppressed
due to a galaxy-ICM interaction \cite{h05}. However, the details of
the physics of such an interaction remain unclear. ACS morphological
information shows that while most of the star-forming galaxies are
disk or irregular galaxies, there exists a population of compact
galaxies with emission ([OII] $\lambda$3727) lines (\cite{h05}; Rosati
et al., in prep.). In RXJ0152, a few star-forming galaxies are as red
($r-i \sim$ 1.2; $i-z \sim$ 0.65) as massive cluster elliptical
galaxies \cite{drl05,h05}, suggesting that these objects (composed of
a mixture of old and young stellar populations) may be in a transition
stage between late-type and S0 galaxies.

Luminous early-type galaxies are observed to dominate the core of many
of these clusters. In some cases, a pair of massive cluster core
elliptical galaxies, located close to the X-ray emission centroid, are
separated by only a few arcseconds (e.g., \cite{b03}). These galaxies
are possibly in a pre-merger phase conducive to the formation of a
massive cD galaxy. However, RXJ0910 lacks of cD galaxies and any clear
evidence of mergers leading to the formation of one ($z \simeq 1.10$;
Mei et al., ApJ, submitted). The analysis of the photometric
properties of galaxies in the ACS intermediate redshift clusters show
that a tight Color-Magnitude Relation (CMR) is already in place at
redshift $z = 1.24$ \cite{b03}. The slope and scatter of the different
CMRs are consistent with passive evolution of early-type galaxies with
the bulk of their stars being formed at redshifts $z > 2$. Yet, the
spectroscopic data available on the cluster RDCS J1252 ($z = 1.24$;
\cite{r04}) suggest that most luminous early-type galaxies in this
cluster host young (post-starburst) stellar populations \cite{r04b}
produced in recent episodes of star formation at $z \simless 2$. A
more detailed presentation and an in depth discussion on the CMRs of
the ACS clusters is given by Mei et al. (this conference). Different
model spectral energy distributions (SED) have also been used to fit
the observed colors of cluster early-type galaxies in order to
estimate galaxy ages and stellar masses. These results are presented
by Rettura et al. (this conference). An extensive morphological
classification of galaxies (about 4700) in the ACS cluster sample has
been carried out by Postman et al. (2005; \cite{p05}). The aim of this
effort is to extend previous studies on the evolution of the
morphology-density relation from the low redshift universe up to
redshift unity. Interesting evolutionary trends are observed
suggesting a transformation of late-type galaxies into S0 galaxies
from $z \sim 1$ down to $z = 0$ (see \cite{p05} and Mei et al. [this
conference]).

\section{Proto-clusters of galaxies}

Distant radio galaxies appear to be good tracers of high density
regions in the universe. This is based on the detection of confirmed
Ly-$\alpha$ emitter and Lyman break galaxy (LBG) overdensities around
radio galaxies at $z > 2$. These candidate overdensities, also called
proto-clusters, have also been identified around lower redshift ($z
\sim 1.6$) radio galaxies \cite{blm03}. Our ACS observations
concentrated on the fields around three radio galaxies: MRC1138-262
($z = 2.16$), TNJ1338-1942 ($z = 4.11$) and TNJ0924-2201 ($z =
5.19$). In the case of TNJ1338, the radio galaxy is surrounded by 12
Ly- emitters \cite{v02} and an overdensity of $\sim$50 LBGs (Overzier
et al., in prep.) with a significance greater than 4$\sigma$. The
angular distribution of these objects is highly filamentary with more
than half of the objects clustered in a 4.4 arcmin$^2$ region whose
center is occupied by the radio galaxy. The latter itself is a very
interesting object. It is the dominant galaxy in the proto-cluster, in
terms of size (about 16 kpc in extension) and luminosity, and it is
surrounded by an extended ($\sim$ 100 kpc), asymmetric Ly- halo. The
ACS optical (rest-frame UV) morphology is characterized by a number of
substructures and the observed flux is likely dominated by forming
stars. The estimated star-formation rate for the whole radio galaxy is
then $\sim$200 M$_{\odot}$ yr$^{-1}$ (see \cite{z05} for
details). These observations strongly suggest that TNJ1338 is destined
to evolve into the brightest proto-cluster galaxy and can likely be,
therefore, a progenitor of the massive early-type galaxies we observe
in local clusters. Likewise, 6 Ly-$\alpha$ emitters have been
spectroscopically confirmed to be around TNJ0924 \cite{v04}, the most
distant radio galaxy known to date \cite{vbdbs99}. Our ACS
observations have allowed us to discover 23 LBGs (V$_{606}$-dropouts
at the radio galaxy redshift) around TNJ0924 (Overzier et al., ApJ,
submitted). This population, which may be confirmed by future
spectroscopic follow-up, may be part of a massive structure in
formation already at $z = 5.2$. Compared to TNJ1338, TNJ0924 is about
three times smaller and its star formation rate is about 20 times
lower (see Overzier et al. [ApJ, submitted] for details).

\section{Conclusions} 

ACS observations are delivering one of the most impressive views of
the formation and evolution of structures in the universe since
$\sim$1 Gyr after the Big Bang. Radio galaxies seem to be excellent
tracers of forming structures (proto-cluters) in the early ($z > 2$)
universe. ACS observations of one of these distant radio galaxies
suggest that these objects may be the progenitors of massive
early-type galaxies in lower redshift clusters. Giant elliptical
galaxies would be formed in major mergers at $z > 2$, with the
associated starburst winds producing the ICM. These luminous
elliptical galaxies are observed to be anchored in place at $z \sim 1$
in the core of rich galaxy clusters, and their photometric properties
indicate a passive evolution of most of their stellar content since
their last episode of major star-formation. Late-type galaxies, on the
other hand, dominate the outer regions of clusters. There is evidence
suggesting that spiral/irregular galaxies in the higer density
environment of clusters would be transformed into earlier type
galaxies, like S0s, however, the physical mechanisms behind that are
still poorly understood. More detailed studies of the evolution of
spiral galaxies in clusters are still required. The comparison between
such studies and those of spiral galaxies in the field (e.g.,
\cite{hfe05}; Hammer et al., this conference), together with the
present knowledge on early-type galaxies, would provide a
comprehensive picture of galaxy evolution as a whole through cosmic
time. Finally, the disturbed distribution and kinematics of cluster
galaxies, ICM and DM show that massive clusters are assembled in a
hierarchical fashion through mergers of multiple clumps since $z
\simgreat 1$. However, more observations are needed in order to bridge
the evolutionary gap between the most distant galaxy clusters known
and the proto-clusters under study.

\acknowledgements{ ACS was developed under NASA contract
NAS5-32865. We thank our fellow ACS team members for their important
contributions to this research and we also thank the support from ESO
staff in Chile and Germany. We are grateful to Ken Anderson, John
McCann, Sharon Busching, Alex Framarini, Sharon Barkhouser, and Terry
Allen for their invaluable contributions to the ACS project at
JHU. This investigation has been partially supported by NASA grant
NAG5-7697.  }

\vfill 
\end{document}